\documentclass{edp-jp4}
\usepackage{graphicx}
\usepackage{amssymb}
\begin{document}


\title{Collective decision making and paradoxical games.}

\author{Juan M.R. Parrondo}\address{Grupo Interdicisplinar de Sistemas
Complejos (GISC) and Dep. F\'{\i}sica At\'{o}mica, Molecular y
Nuclear. Universidad Complutense de Madrid, 28040-Madrid, Spain.}
\author{Luis Dinis}\sameaddress{1}
\author{Eduardo Garc\'{\i}a-Tora\~no}\sameaddress{1}
\author{Bel\'{e}n Sotillos}\sameaddress{1}

\maketitle

\begin{abstract}
We study an ensemble of individuals playing the two games of the
so-called Parrondo paradox. In our study, players are allowed to
choose the game to be played {\em by the whole ensemble} in each
turn. The choice cannot conform to the preferences of all the
players and, consequently, they face a simple frustration phenomenon
that requires some strategy to make a collective decision. We
consider several such strategies and analyze how fluctuations can be
used to improve the performance of the system.
\end{abstract}


\section{Introduction}

Collective decision making is a major topic not only in economics
and public choice theory \cite{nitzan}, but also in communication,
computer science, machine learning, game theory, and control theory.
 The so-called paradoxical games or Parrondo paradox
\cite{nature,contphys} have provided a simple example of collective
decision making exhibiting interesting features, such as the
inefficiency of voting \cite{physicaA} and of short-term
optimization \cite{epl}.

The paradox was initially designed as an illustration of Brownian
ratchets. However, its simplicity has helped to explore possible
applications of the ratchet effect to quantum games \cite{qg},
quantum decoherence \cite{qdec}, microbial evolution
\cite{hanta,wolf}, deterministic chaos \cite{caos}, and stochastic
control and optimization \cite{prl,epl2}.

The paradox consists of two losing gambling games, A and B, which
yield a winning game when alternated either in a periodic or random
way. In the original formulation of the paradox, an individual plays
against the casino with the following rules. In game A, the player
has a probability $p_A=1/2-\epsilon$ of winning one euro and a
probability $1-p_A=1/2+\epsilon$ of losing one euro, where the bias
$\epsilon$ is a small positive number. This game can be considered
as a bet of one euro on the toss of a slightly biased coin.

\begin{figure}
\begin{center}
\includegraphics[width=0.6\textwidth]{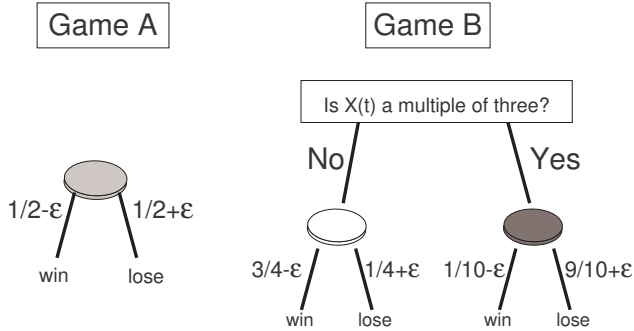}
\caption{  \label{figrules} Rules of the paradoxical games.}
\end{center}
\end{figure}

The second game, B, is played with two biased coins, a ``bad coin''
and a ``good coin''. The player must toss the bad coin if her
capital $X(t)$ is a multiple of $3$, the probability of winning
being $p_{\rm bad}=1/10-\epsilon$. Otherwise, the good coin  is
tossed and the probability of winning is
 $p_{\rm good}=3/4-\epsilon$. The rules of games A and B are
 represented in Fig.~\ref{figrules}, in which  darkness
indicates the ``badness" of each coin.

The rules are such that the two games are losing if $\epsilon>0$,
i.e., the mean value of the capital $X(t)$ is a decreasing function
of the number $t$ of turns played. However, for many random and
periodic sequences, such as ABBABBABB..., the capital increases on
average.

Recently, the possible relevance of the paradox has been questioned
by arguing that it becomes trivial if the player is allowed to
choose the game in each turn \cite{iyengar}. This argument is itself
quite trivial. The choice is very simple: if $X(t)$ is a multiple of
three, one has to choose game A to avoid the use of the bad coin in
game B; on the other hand, if $X(t)$ is not a multiple of three,
then the best choice is B. With this strategy, the capital is
clearly increasing on average. Moreover, this is the optimal
strategy for a single player.

However, if we now consider an ensemble of players who have to make
a {\em collective decision}, i.e., to choose the game to be played
by the whole ensemble, then the problem becomes more interesting.
Some of the players will have a capital multiple of three and will
prefer to play game A, while the rest will choose B. This is a
simple frustration phenomenon, which makes the paradoxical games a
suitable model for analyzing the efficiency of collective decisions.

Some possible strategies have been recently studied finding new
counter-intuitive results. For instance, a  maximization of the
profits of a single turn (short-term optimization) yields worse
results than a ``blind'' periodic strategy \cite{epl}. Moreover, a
``democratic'' decision, which chooses the game preferred by the
majority  of players, can be losing whereas a random choice is
winning \cite{physicaA}. Part of these results have inspired the
consideration of Brownian ratchets as feedback control systems
\cite{prl}.

In this paper, we review the voting and the short-term optimization
strategies as special cases of a democratic choice with variable
threshold. Then we consider decisions taken by a subset of players
---``dictators'' and ``oligarchs''---, although affecting the whole
ensemble. These new strategies or decision protocols are of interest
for the general theory of stochastic control and decision making in
random environments, and can also be translated to controlled
Brownian ratchets, in order to improve its performance or induce
segregation of target particles \cite{linkeprivate}.

The paper is organized as follows. In section \ref{democracy} we
review the performance of the democratic decision with different
thresholds. In section \ref{dictator} we consider decisions taken by
a single player or dictator, a model which is analytically solved in
the Appendix. In section \ref{oligarchy}, we analyze a strategy
where a reduced set of players, the oligarchs, choose the game to be
played in each turn. Finally, we  present our main conclusions in
section \ref{conclusions}.

\section{Democracy}
 \label{democracy}

We consider along the paper a set of $N$ individuals playing
independently against a casino either game A or B with $\epsilon=0$.
Each player votes for A or B according to her own interest, i.e.,
those players with capital multiple of three vote for A and the rest
for B. Finally, game A is chosen in turn $t$ if the fraction
$\pi_0(t)$ of votes for A is above a certain threshold $u\in [0,1]$.

In this model, some analytical results are known for $N\to\infty$.
Let us define the state of the system at turn $t$ as the vector:
\begin{equation}\label{state}
\vec{\pi}(t)=\left(\begin{array}{c} \pi_0(t) \\ \pi_1(t) \\ \pi_2(t)
\end{array}\right)
\end{equation}
where $\pi_i(t)$ is the fraction of players whose capital $X_k(t)$
can be written as $X_k(t)=3n+i$ for some integer $n$
\cite{contphys}. For $N\to\infty$, the evolution of this state is
deterministic and follows a piecewise linear recurrence map, which
depends on the threshold $u$. Cleuren \cite{tesisbart} and Dinis
\cite{tesisluis} have studied the attractors of such maps and the
sequence of selected games corresponding to each attractor.

\begin{figure}
\begin{center}
\includegraphics[width=0.49\textwidth,clip=true]{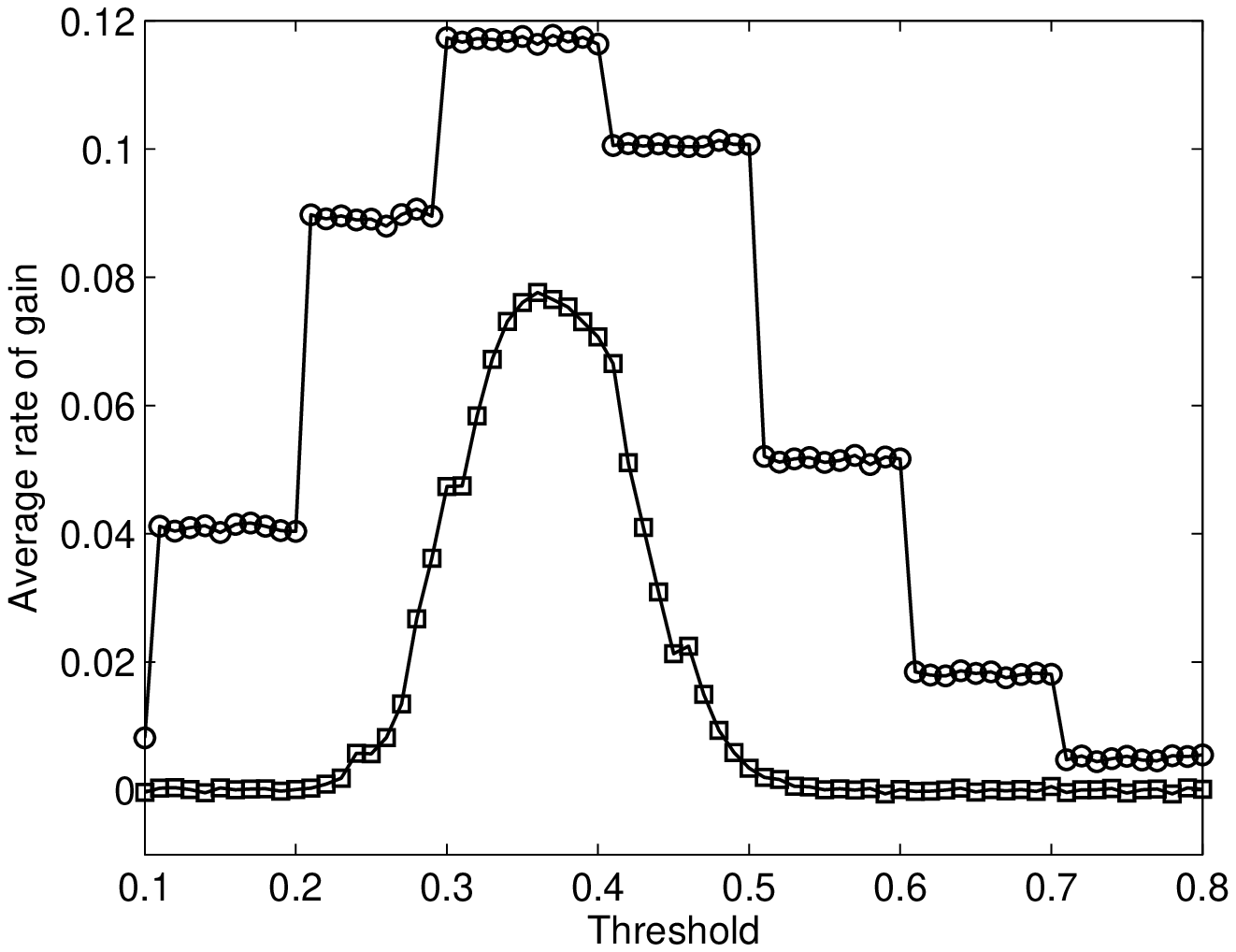}
\includegraphics[width=0.49\textwidth,clip=true]{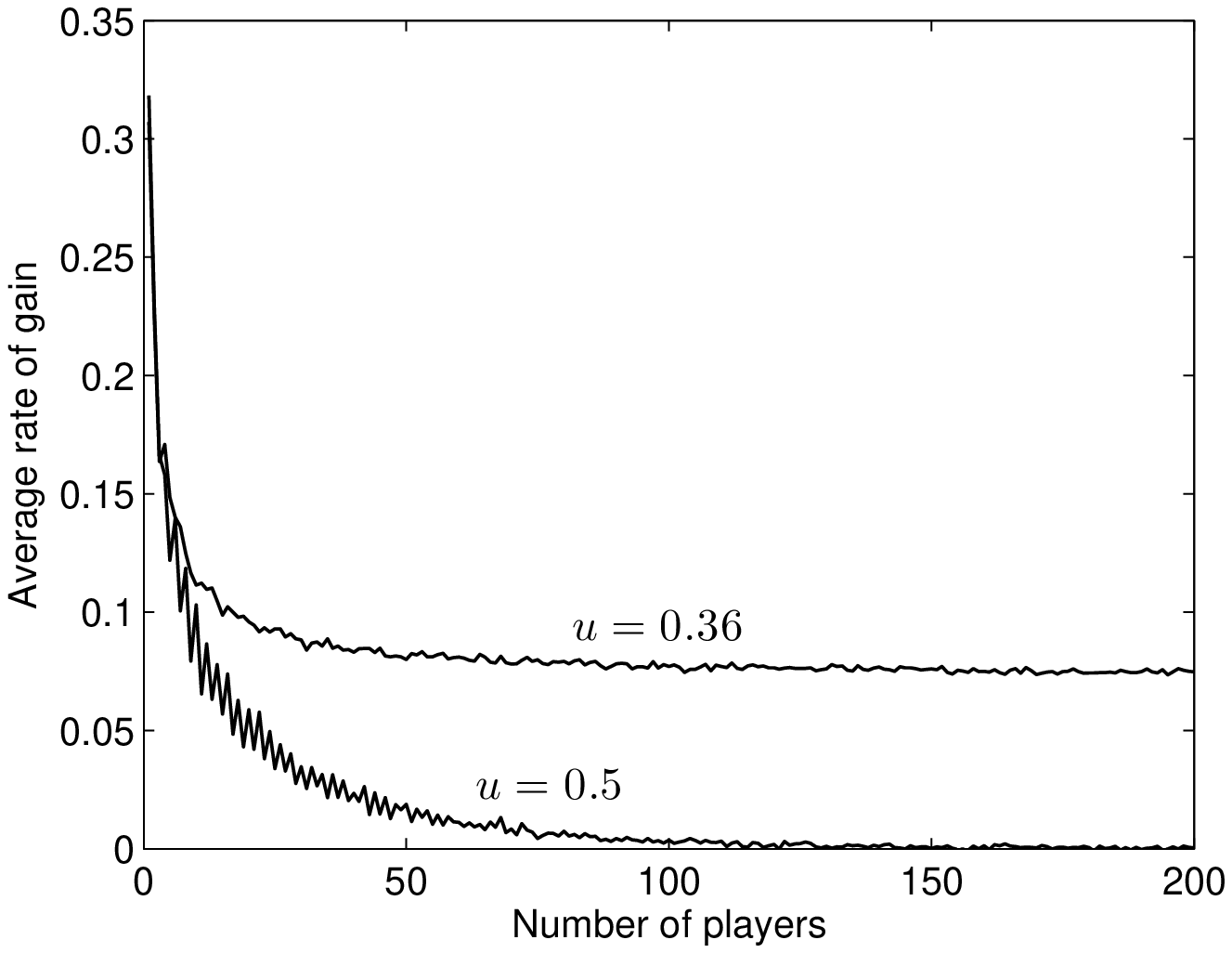}
\caption{\label{demoN}  {\em Left:} Average rate of gain per player
as a function of the threshold, i.e., minimal fraction of votes for
A to select this game, for $N=10$ (circles) and $N=100$ (squares)
players. {\em Right:} Average rate of gain per player as a function
of the number of players for two values of the threshold $u$: simple
majority, $u=1/2$, and optimal threshold for $N=100$, $u=0.36$.}
\end{center}
\end{figure}

In the case of simple majority, $u=1/2$, it is not difficult to see
that the stationary sequence of games is BBB.... The reason is that,
if B is played a large number of turns in a row, $\pi_0(t)$ tends to
$5/13$ \cite{contphys} which is far below the threshold 1/2. Since B
is a fair game (recall that we have set $\epsilon=0$), the
stationary gain for this simple majority rule vanishes in the limit
$N\to\infty$ (cfr. Fig. \ref{demoN}).

Another important case is $u=5/13$, since it corresponds to a
short-term maximization of returns \cite{epl}. In this case, one can
show \cite{tesisbart,tesisluis} that the sequence of games in the
stationary regime is ABBABB... yielding a gain per player
$g_{short}=\frac{2416}{35601}=0.06789...$.

However, this short-term maximization does not give rise to the
actual optimal sequence, which is  ABABBABABB... \cite{tesisluis}.
Such a sequence can be achieved in the limit $N\to\infty$ only by
using a threshold within a narrow interval
$u\in(1/3,11945/35601)\approx(0.3333,0.33550)$ \cite{tesisbart}.
This optimal sequence yields an average gain per player and turn
$g_{opt}=\frac{514608}{47747645}=0.07568...$

For finite ensembles the results are very similar. In Fig.
\ref{demoN} (left) we show the average rate of gain per player as a
function of the threshold, for $N=10$ and $N=100$ players. We see
that the simple majority rule ($u=1/2$) yields a small gain even for
$N=10$. On the other hand, a threshold around $u=0.36$ turns out to
be optimal for both small ($N=10$) and large ($N=100$) ensembles.
Notice however that $u=0.36$ is no longer optimal for infinite
ensembles, since gives rise to the stationary periodic sequence
ABBABB... For $N=100$, fluctuations are still important (of the
order of 10\%), and the threshold $u=0.36$ makes a better use of
such fluctuations yielding an average gain $g\approx 0.776$ (recall
that the average gain with the optimal periodic sequence ABABB... is
$g_{opt}\approx 0.07568$). The associated sequence of games is
similar to ABBABB... plus some random ``mutations'' that depend on
the random state of the system and improve its overall performance.

Finally, we show in Fig. \ref{demoN} (right) the average rate of
gain per player as a function of the number of individuals in the
ensemble for the simple majority threshold ($u=1/2$) and for the
optimal threshold found in the cases $N=10$ and $N=100$, namely,
$u=0.36$. Both thresholds give similar returns for small ensembles
but the simple majority rule rapidly decreases to zero as $N$
increases.

We have to stress that these results are very sensitive to the rules
of the games. For instance, the average gain in the case of the
short-term optimization vanishes for $N\to\infty$ if the probability
to win and loose in each game do not sum up to one, i.e., if there
is a non zero probability  that players do not change their capitals
in each turn \cite{epl}.

\section{Dictatorship}
\label{dictator}

\begin{figure}
\begin{center}
\includegraphics[width=.49\textwidth]{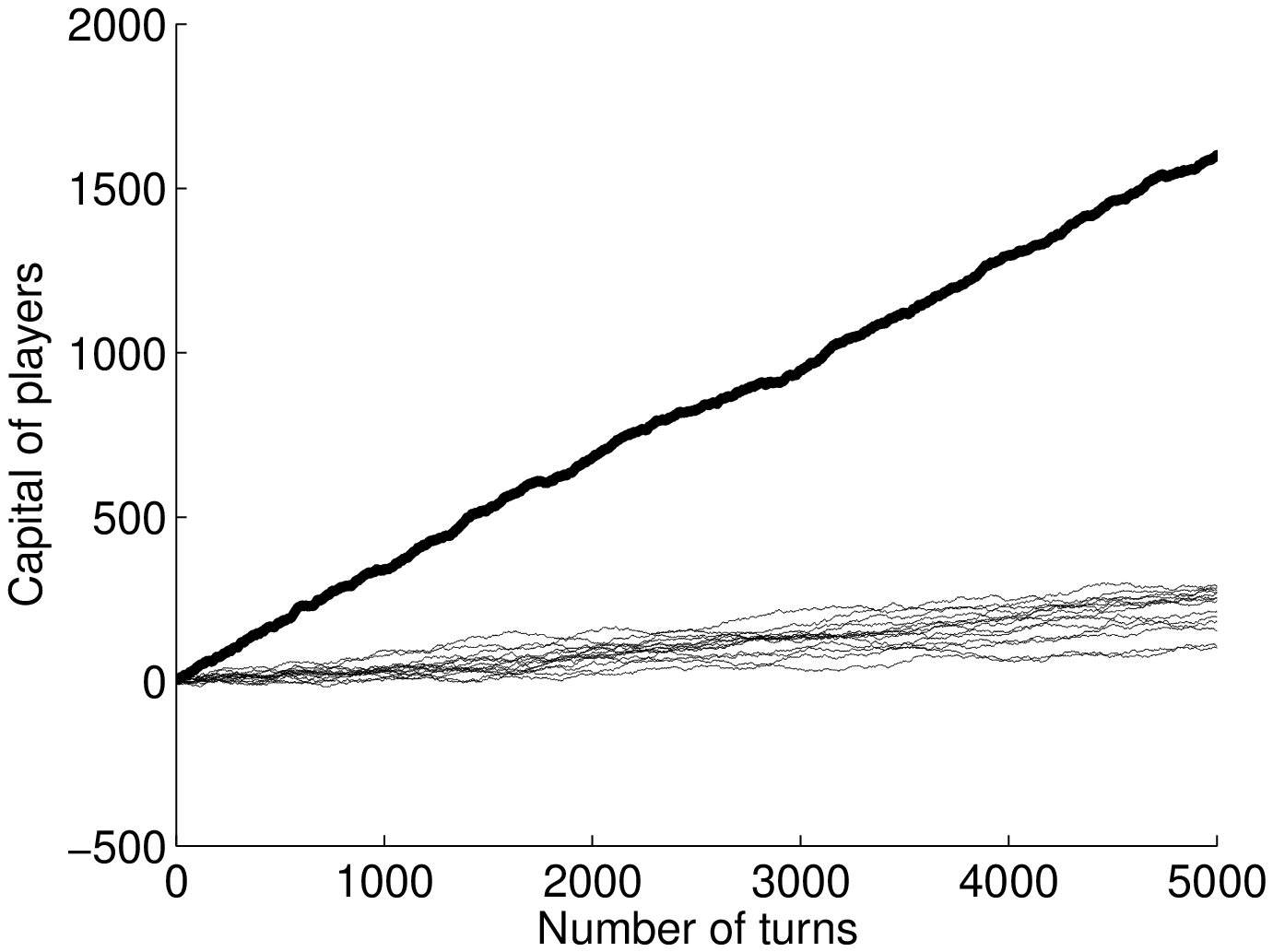}
\includegraphics[width=.49\textwidth]{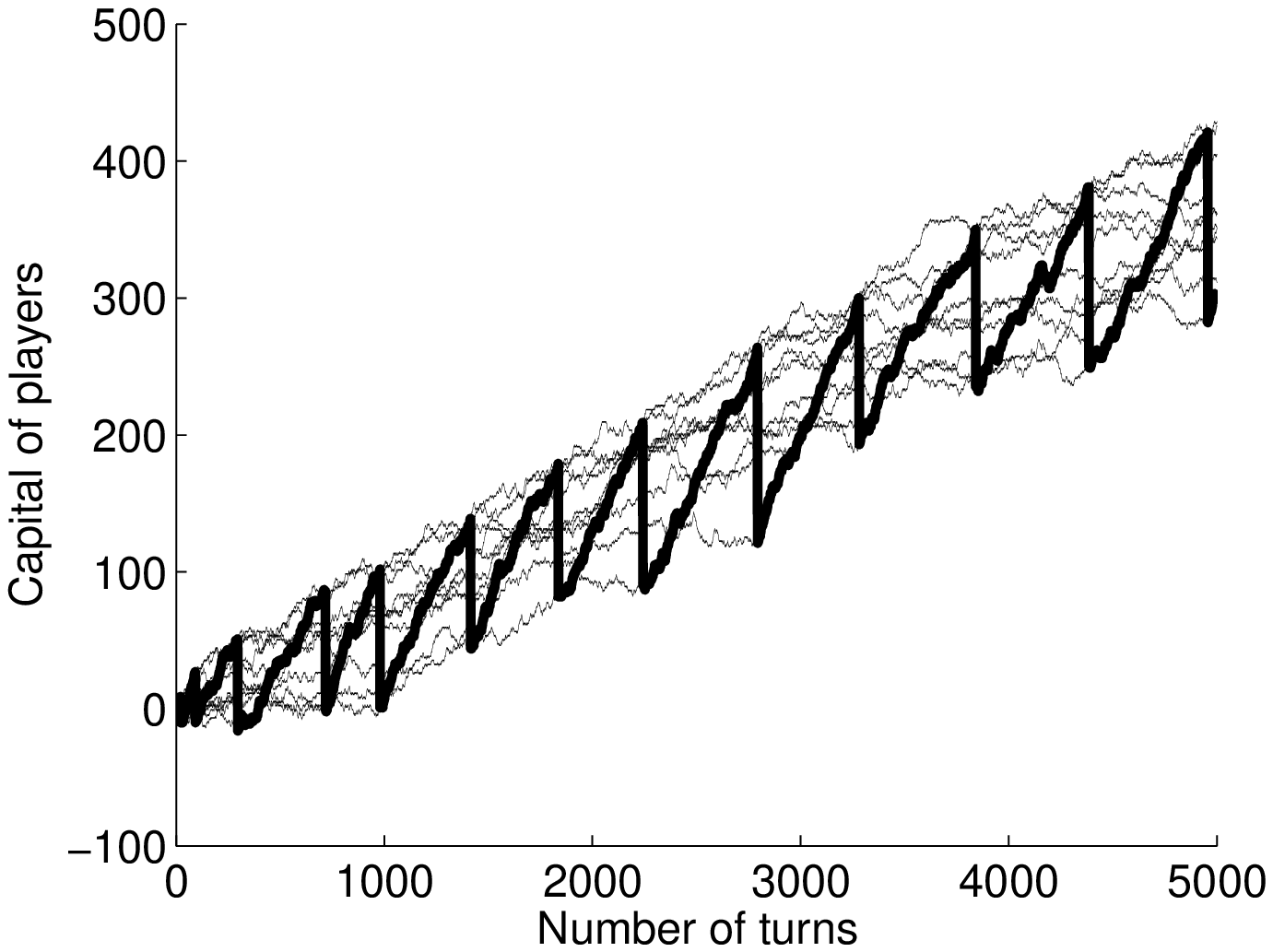}
\caption{\label{figdic}  Simulation results for the fixed ({\em
left}) and alternating ({\em right}) dictator model,  with 10
players (9 citizens and one dictator). The capital of the dictator
at each turn is indicated by the thick line.}
\end{center}
\end{figure}

 We now consider a model where the choice of the game is made
by a single player, the dictator, whereas the rest of the players
---the ``citizens''--- must accept her decision. Hence, the whole
ensemble plays game A if the dictator's capital is a multiple of
three and B otherwise. The results of a simulation with 10 players
(the dictator plus 9 ``citizens''), is shown in Figure \ref{figdic}
(left).

This model can be solved analytically, as explained in the Appendix,
and the average gain per turn and player for both the dictator and
the citizens read, respectively:
\begin{eqnarray}\label{slopes}
g_d &=& \frac{12}{37}  =  0.324... \nonumber \\
g_c &=&  \frac{628224}{13685449} = 0.0459...
\end{eqnarray}
which  coincide with those observed in simulations (cfr. Fig.
\ref{figdic}, left).

As expected,  dictator's earnings grow fast since she is playing the
optimal strategy for a single player. It is not so clear at first
sight what will happen to the rest of the players or citizens. Note
that the choice of the game does not depend on their capital at all
and therefore it should behave exactly as if a given sequence of
games was imposed. This sequence is the result of the random walk
performed by the dictator's capital, so it is a random sequence.
Interestingly enough, the capital of the citizens increases at a
bigger rate than in the optimal random and uncorrelated sequence of
games. This optimal random sequence  is achieved when games A and B
are played at frequencies $\gamma$ and $1-\gamma$, respectively,
with $\gamma\approx 0.4146$, and the resulting maximum gain per turn
is $g_{max,rand}\approx 0.02620$ \cite{chaos}, significantly smaller
than $g_c\approx 0.0459$ in Eq. (\ref{slopes}). This means that the
random sequence of games imposed by the dictator exhibits time
correlations which are beneficial for the rest of the players,
although the sequence is not correlated with their own state.

\begin{figure}[h]
\begin{center}
\includegraphics[width=.5\textwidth]{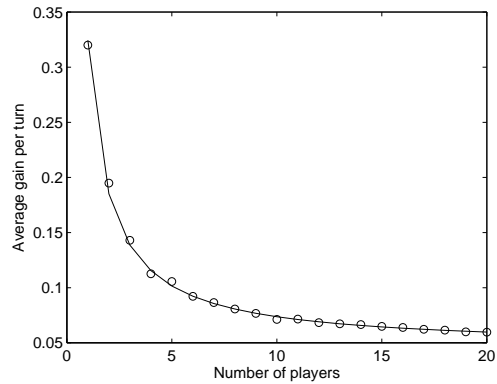}
 \caption{\label{rotN} Average gain per turn of the whole community
for the fixed or cyclic dictator (solid line, analytical results
from Eq. (\ref{averagealt})), and for the alternating dictator
(circles, simulations over $10^5$ turns).}
\end{center}
\end{figure}

The performance of the whole community under the dictator rule is
given by the average gain per turn and player:
\begin{equation}\label{averagealt}
g=\frac{g_d+(N-1)g_{c}}{N}
\end{equation}
This average gain $g$ is shown in Fig. \ref{rotN} (solid line). The
dictator protocol is better than the democracy with a simple
majority rule, but worse than democracy with the optimal threshold.
In this protocol, as clearly seen in Fig. \ref{figdic} (left), the
dictator separates from the rest of the players. A similar strategy
in ratchets can be useful for segregation of target particles
\cite{linkeprivate}.

One could think that the high rate of gain of the dictator will
further help to boost the capital of the whole community if the role
of the dictator is alternated among all the players. We have
explored several versions of this idea. In first place, a purely
cyclic dictator among $N$ players yields the same average rate of
gain $g$ given by (\ref{averagealt}), which clearly converges to
$g_c$ for large $N$. This average rate of gain is the same as in the
case of a fixed dictator, although, with a cyclic dictator there is
no segregation of a single individual.

 A more elaborated version consists of choosing the poorest
 individual to play the role of the dictator until she becomes the
 richest. The result of this protocol is shown in Fig. \ref{figdic} (right).
 As in the case of cyclic dictators, the dispersion of capitals is small and
 all citizens benefit from the strategy, but in this case even
 those who {\em never} become dictators. However, the
 overall performance of this helping-the-poorest protocol and of the
 purely cyclic dictator are the same, as shown in Fig. \ref{rotN}.
 The reason is that  Eq. (\ref{averagealt}) still holds within time intervals
where the dictator does not change (see Fig. \ref{figdic}, right),
and each of these changes is in fact a relabeling of the players
with no effect on the total capital.

These results indicate that Eq. (\ref{averagealt}) is the optimal
gain if we use only the information of the state of {\em one} of the
players in each turn, although we do not have a rigorous proof of
this statement. We will discuss further on this point in Sec.
\ref{conclusions}.

\section{Oligarchy}
\label{oligarchy}

Since democracy is better for small ensembles, one could try to
improve the performance of the whole community allowing a reduced
subset of the ensemble, or ``oligarchy'', to decide the game to be
played. The oligarchs take their decision by voting with a given
threshold as in the democratic protocols of Sec. \ref{democracy}.

\begin{figure}[h]
\begin{center}
\includegraphics[width=0.49\textwidth,clip=true]{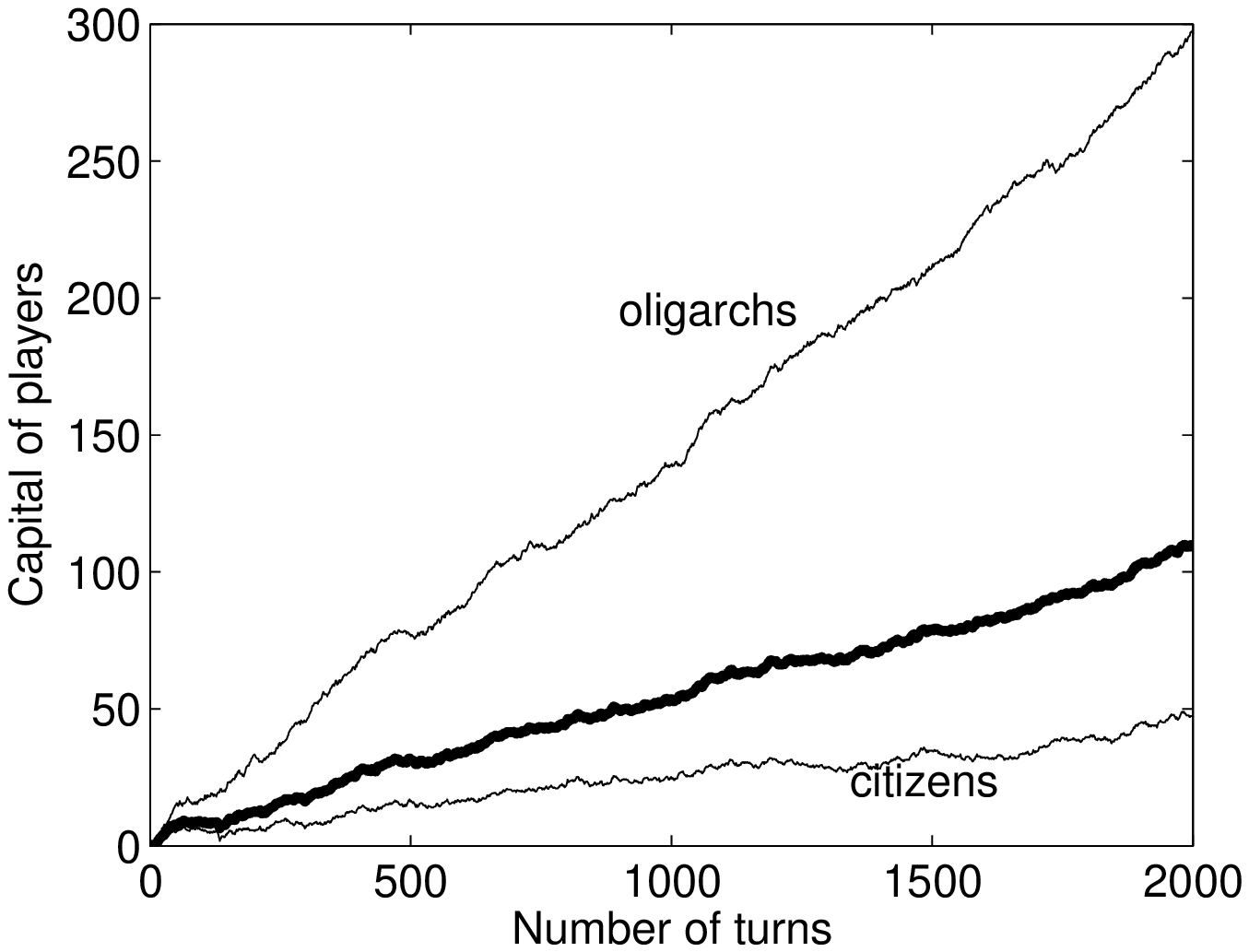}
\includegraphics[width=0.49\textwidth,clip=true]{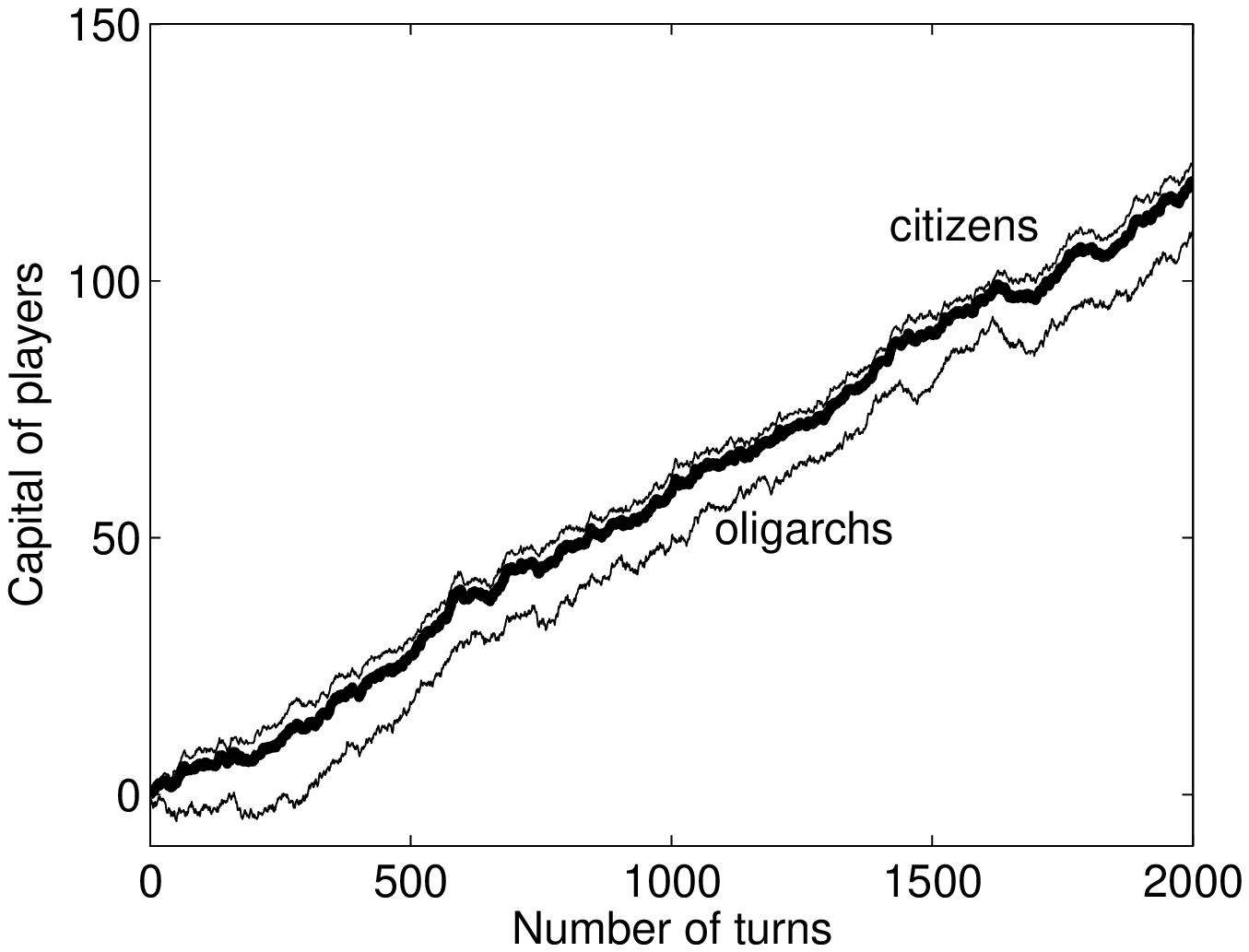}
\caption{ \label{figo}  Evolution of the capital in the oligarchy
protocol. We plot the average gain per player of 5 oligarchs and 10
citizens, with threshold $5/13$. {\em Left:} fixed oligarchy. {\em
Right:} adaptive oligarchy, formed by the 5 poorest players in each
turn.}
\end{center}
\end{figure}

Figure \ref{figo} (left) presents a typical realization of the game
with 5 oligarchs and 10 citizens. The oligarchs use a short-term
optimization with threshold $u=5/13$. Only the average rate of gain
for oligarchs and citizens are shown, for clarity. The capital of
the oligarchy behaves exactly as in the democracy case with only 5
players. As in the dictatorship, the capital of the rest of the
citizens also increases at a larger rate than in any random
combination of games.

\begin{figure}[h]
\begin{center}
\includegraphics[width=0.49\textwidth,clip=true]{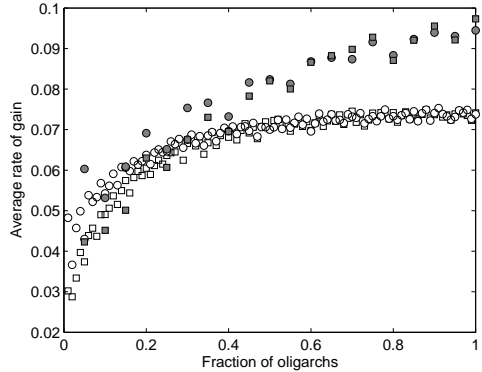}
 \caption{ \label{figoN} Rate of gain per player in the oligarchy
protocol as a function of the fraction of oligarchs with threshold
$u=0.36$: fixed oligarchy (circles) and adaptive (squares); for
$N=20$ (grey symbols) and $N=100$ (white symbols).}
\end{center}
\end{figure}

As we have seen in section \ref{democracy}, the rate of gain of a
set of players taking a democratic decision decreases with the
number of players. Consequently, one could ask if there exists an
optimal size for the oligarchy. Our simulations, shown in  Fig.
\ref{figoN}, indicate that this optimal oligarchy is the whole
ensemble of players. The reason is that increasing the size of the
oligarchy has two effects: a) the oligarchs' weight in the total
average increases and b) the sequence of games is closer to ABB (or
ABABB, depending on the threshold) which is sound for everybody.

Again, one could try to improve the average returns by allowing the
poorest players to choose the game. With this idea in mind, we have
analyzed a special case of {\em adaptive} oligarchy: in each turn,
the oligarchy is formed by the $n_{olig}$ poorest players. The
results are shown in both Figs. \ref{figo} and \ref{figoN}. As in
the case of the alternating dictator, giving the role of oligarchs
to the poorest players has the only effect of keeping closer all
individuals in the ensemble, but does not affect the total capital:
Fig. \ref{figoN} shows that the total capital of the ensemble in the
fixed or adaptive oligarchy are very similar, both for small
($N=20$) and large ($N=100$) ensembles. In this latter case, the
result was expected, since for large oligarchies, no matter if they
are fixed or not, the sequence of games is close to ABBABB... plus
some small fluctuations beneficial only for the oligarchs, but not
affecting the rest of players.

\section{Conclusions}
\label{conclusions}

We have studied several strategies for a collective decision making
in ensembles of individuals playing paradoxical games against the
casino.

The optimal ``blind'' choice of games is the periodic sequence
ABABBABABB... For small ensembles a democratic choice with a
threshold around 0.36 for selecting game A can do better than the
optimal ``blind'' sequence of games, even up to $N=100$ players.
However, this is expected since, for this size, fluctuations are of
the order of 10\%, big enough to be used by strategies depending on
the actual state of the system.

The present study prompts several questions about how fluctuations
can be used to improve the performance of the ensemble. Firstly, our
results with dictators and oligarchies indicate that the gain that
can be achieved depends more on the information used in each
strategy than on its details. For instance, our simulations show
that adaptive and fixed dictators yield the same gain per player.
Oligarchies exhibit a similar feature. Secondly, in the case of
oligarchy, we have found that it is always better the decision taken
by the whole ensemble (at least for certain thresholds close to the
optimal one) than by any subset. That is, it is better to use the
maximum information available.

It is worth to mention that the actual optimal strategies for $N$
finite are based on the complete state of the system, i.e., on the
vector (\ref{state}) \cite{tesisbart,tesisluis}, whereas the only
information that we have used here is the fraction $\pi_0(t)$ of
players in the bad coin of game B. However, the analysis could be
extended to incorporate this additional information without
affecting our main conclusions.

Touchette and Lloyd \cite{lloyd} have explored the limitations that
thermodynamics imposes to the energetics of some controlled physical
systems, based on the information used in the feedback control. In
this paper we have shown that analogous limitations apply to the
gain of players or, equivalently, to the velocity of random walkers
in non-homogeneous and asymmetric media. We believe that our results
could be explained and generalized by a theoretical framework that
relates the measure of the information used in a control protocol
with the performance of such a protocol in stochastic systems with
feedback, following similar lines as those in Ref. \cite{lloyd}.

\section*{Acknowledgements}
 This work is financially supported by Spanish
Grant FIS2004-271 (MCyT, Spain) and Grant SANTANDER/COMPUTENSE
PR27/05-13923. B.S. and E.G.-T. also acknowledge financial support
from the Comunidad Aut\'{o}noma de Madrid (Becas de Excelencia).

\Appendix

\section{Analytical solution of the fixed-dictator protocol}

Let $X_d(t)$ and $X(t)$ be the capital of the dictator and of a
given citizen at time $t$, respectively. We define the vector ${\bf
Y}(t)$ as:
\begin{equation}
{\textbf{Y}}(t)= \left(\begin{array}{cc} X_d(t)\ \textrm{mod}\ 3 \\
X(t)\ \textrm{mod}\  3\end{array}\right)
\end{equation}
which can take on nine different values: (00), (01), (02), (10),
(11), (12), (20), (21, (22). The vector ${\bf Y}(t)$ is a Markov
chain with the following transition matrix (we number the nine
states in the same order as in the previous list):
\begin{equation} \Pi =
 \left( \begin{array}{ccccccccc} 0 & 0 & 0 & 0 &
\frac{1}{16} & \frac{3}{16} & 0 & \frac{3}{16} & \frac{9}{16}
\\ 0 & 0& 0& \frac{1}{40} & 0 & \frac{1}{16} &
\frac{3}{40} & 0 & \frac{3}{16}
\\ 0 & 0 & 0 & \frac{9}{40} & \frac{3}{16} & 0 &  \frac{27}{40} &
\frac{9}{16} & 0
\\ 0 & \frac{1}{4} & \frac{1}{4} & 0 & 0 & 0 & 0 & \frac{1}{16} &
\frac{3}{16} \\ \frac{1}{4} & 0 & \frac{1}{4} & 0 & 0 & 0 &
\frac{1}{40} & 0 & \frac{1}{16} \\ \frac{1}{4} & \frac{1}{4}
& 0 & 0 & 0 & 0 & \frac{9}{40} & \frac{3}{16} & 0 \\
0 & \frac{1}{4} &
\frac{1}{4} & 0 & \frac{3}{16} & \frac{9}{16} & 0 & 0 & 0 \\
\frac{1}{4} & 0 & \frac{1}{4} & \frac{3}{40} & 0 & \frac{3}{16} & 0
&  0 & 0 \\ \frac{1}{4} & \frac{1}{4} & 0 & \frac{27}{40} &
\frac{9}{16} & 0 & 0 & 0 & 0
\end{array}\right)
\end{equation}
Each entry of the matrix is the transition probability from a state
to another, and can be derived directly from the rules of the game.
For instance, the transition probability from (01) to (12) is
$p_a(1-p_a)=1/4$, since the dictator in the initial state has a
capital multiple of three and chooses A as the game to be played.
This corresponds to $\Pi_{6,2}$. The transition probability from
(21) to (02) is $p_{b2}p_{b1}=9/16$, since in this transition both
the dictator and the citizen win playing game B. Therefore
$\Pi_{3,8}=9/16$.

The matrix $\Pi$ is an stochastic matrix with one eigenvalue equal
to 1. The corresponding eigenvector, after normalization, is the
stationary probability distribution of the Markov chain and reads:
\begin{equation}
P_{st}=\frac{\left( 1759355,633035, 2416011, 1246210, 1218764,
1233796, 1684790, 1368644, 2124844\right)}{13685449}
\end{equation}

From this distribution one can calculate the winning probabilities
for the dictator and the citizens, taking into account the winning
probability in each state. For instance, for the dictator we have:
\begin{equation} P_{win,dict}=(P_{st,1}+P_{st,2}+P_{st,3})p_a+
(P_{st,4}+P_{st,5}+P_{st,6}+P_{st,7}+P_{st,8}+P_{st,9})p_{b1}
\end{equation}
since the dictator wins with probability $p_a$ in the three first
states and with probability $p_{b1}$ in the other six states.

Finally, the average rate of gain per player and per turn is,
respectively:
\begin{eqnarray}
g_d &=&\lim_{t\to\infty} \frac{X_d(t)}{t} = 2P_{win,dict}-1 =\frac{12}{37}  =  0.324... \nonumber \\
g_c&=&\lim_{t\to\infty} \frac{X(t)}{t} = 2P_{win,cit}-1
=\frac{628224}{13685449} = 0.0459...
\end{eqnarray}
These values coincide with the results of the simulations for
$\epsilon=0$.


\end{document}